\documentclass[aps,amsmath,amssymb,nofootinbib,twocolumn,pr]{revtex4-1}%
\usepackage{graphicx}
\usepackage{color}
\usepackage{verbatim} 
\usepackage{color}
\usepackage{bbold}
\usepackage{dcolumn}
\usepackage{bm}
\usepackage{upgreek}
\usepackage[unicode=true,colorlinks=true,citecolor=blue,urlcolor=blue]{hyperref}
\usepackage{ulem}
\usepackage{physics}

\newcommand*\xoverline[2][0.75]

\begin{document}

\title{Sommerfeld enhancement factor in two-dimensional Dirac materials}

\author{N.~V.~Leppenen}
\author{L.~E.~Golub} 	
\author{E.~L.~Ivchenko} 	
\affiliation{Ioffe Institute, 194021 St. Petersburg, Russia}


\begin{abstract}
In this work the above-band gap absorption spectrum in two-dimensional Dirac materials is calculated with account for the interaction between the photocarriers. Both the screened Rytova--Keldysh and pure Coulomb attraction potentials between the electron and hole are used in the study. We find that, in the materials under consideration, the interaction enhances the absorbance in the narrow interband edge region, in a sharp contrast to the band model with the parabolic free-carrier energy dispersion. We develop an approximation of the weak interaction which allows us to reproduce the main features of the exactly calculated Sommerfeld factor. We show a substantial reduction of this factor at higher frequencies due to the single-particle energy renormalization.
\end{abstract}

\maketitle

\section{Introduction} Effects of interaction between the electrons and holes created by the absorption of photons in semiconductors are under study 
starting from 1950s up to now~\cite{Gross1952,Gross1956,Excitons,Glutsch,Bayer,Shubina}. 
In an undoped semiconductor crystal, the interband light absorption spectrum consists of a series of discrete levels of bound exciton states below the bandgap edge $E_g$ and a continuum due to unbound electron-hole pairs above the band edge. However, a residual effect of the Coulomb mutual attraction of an electron and a hole gives rise to a correlation of their relative position in space and an enhanced wavefunction overlap. This influences the optical transition matrix and affects the spectral shape of the continuum absorption at $\hbar\omega>E_g$ where $\omega$ is the frequency of light. The ratio of the absorption coefficients above the band edge with and without the Coulomb interaction is called the Sommerfeld factor or the Coulomb enhancement factor. In a three-dimensional (3D) semiconductor, due to this factor the absorption coefficient changes the square-root dependence $(\hbar \omega - E_g)^{1/2}$ to a constant absorption closely above the bandgap, as shown in the seminal paper of Elliott \cite{Elliott}. In a two-dimensional (2D) system, the bare band-to-band step absorption spectrum is multiplied by the 2D Sommerfeld factor which equals 2 at the gap edge and unity far above the gap \cite{Sugano1966,Miller1984}. In both 3D materials and 2D semiconductors, III-V and II-VI based quantum wells, the changes in the absorption spectrum are seen in the range of several exciton Rydberg energies~\cite{EL_book,Haug_Koch_book}.

The classical theories of the exciton oscillator strength~\cite{Elliott,Sugano1966,Miller1984,Haug_Koch_book,EL_book,EL_book} are valid provided the exciton Rydberg is small compared to the band gap and the exciton wave function is just a linear combination of products of free electron states in the conduction band and free hole states in the valence band. Nowadays, the research interest is focused on excitons in the 2D Dirac materials, with the charge carriers being well described by the 2D Dirac equation and the electron-hole interaction energy being less but comparable to the energy gap~$E_g$~\cite{rev1,rev2}. In this case, the exciton wave function is described by four components $\psi_{\lambda_e, \lambda_h}$ ($\lambda_e = \pm, \lambda_h = \pm$)  with the electron and the hole having both signs of energy, including the case when they both are in the conduction band ($\lambda_e = +, \lambda_h = -$) or in the valence band ($\lambda_e = -, \lambda_h = +$)~\cite{Two_body_graphene, Berman_2013, Berman_2016, Exc_trion_TMD, Exc_TMD_energies,Leppenen_2020}. 

In Ref.~\cite{Leppenen_2020}, a theory of the bound-exciton oscillator strength has been developed accounting for the conduction-valence band coupling in the 2D Dirac model and the four-component structure of the exciton wave function. In the present work, we extend the theory \cite{Leppenen_2020} and study how the interaction affects the optical matrix elements and absorption above the band edge of the 2D Dirac materials. Earlier, the study of the effect of nonparabolicity on absorption spectra was carried out in Refs.~\cite{PbS,Exc_top_ins,Somm_fact_TMD}. However, in Ref.~\cite{PbS}, the band-nonparabolicity effect in a lead-sulfide quantum well is taken into account merely by using energy-dependent effective masses of an electron in the conduction band and a hole in the valence band. In Refs.~\cite{Exc_top_ins,Somm_fact_TMD}, the exciton is unjustifiably described by a single variable equivalent to the component $\psi_{++}$ of the exciton wave function and thereby the four-component structure of the latter is lost.

The paper is organized as follows. In Sec.~\ref{Abs_calc}  we derive the working equations and present numerical calculation of the absorbance in the 2D Dirac materials. 
In Sec.~\ref{Pert_theor}, a perturbation theory is developed assuming the electron-hole interaction to be small, a first order correction to the absorbance is found and the results are compared with the exact calculation. 
In Sec.~\ref{Concl} we summarize the results.

\section{Sommerfeld factor calculation}
\label{Abs_calc}
As well as in Ref.~\cite{Leppenen_2020} we consider a two-valley 2D Dirac semiconductor with the valleys $K$ and $K'$ which are  time-reversal of each other. The components $\psi_{++}, \psi_{+-}, \psi_{-+}, \psi_{--}$ of the electron-hole-pair wave functions are conveniently presented as a four-component vector ${\bm \Psi}_{\rm exc}$. For the direct-gap excitations with the electron lying in the $K$ valley and the hole lying in the $K'$ valley, the vector ${\bm \Psi}_{\rm exc}$ satisfies the two-particle Schr\"odinger equation \cite{Leppenen_2020}
\begin{equation} \label{Schr}
{\cal H}_{\rm exc}  {\bm \Psi}_{\rm exc}({\bm \rho}_e, {\bm \rho}_h)  = E {\bm \Psi}_{\rm exc}({\bm \rho}_e, {\bm \rho}_h) 
\end{equation}
with the 4$\times$4 matrix Hamiltonian
\begin{equation} \label{calH}
{\cal H}_{\rm exc} = {\cal H}^K(\hat{\bm k}_e) \otimes \mathbb{1}  +  \mathbb{1} \otimes {\cal H}^{h,K'}(\hat{\bm k}_h)  + V({\bm \rho}) \:.
\end{equation}
Here $\mathbb{1}$ is the 2$\times$2 unit matrix, ${\bm \rho}_e, {\bm \rho}_h$ are the electron and hole 2D space coordinates, ${\bm \rho} = {\bm \rho}_e - {\bm \rho}_h$, $\hat{\bm k}_e = - {\rm i} {\bm \nabla}_e$, $\hat{\bm k}_h = - {\rm i} {\bm \nabla}_h$, $V({\bm \rho})$ is the electron-hole attraction potential [$V({\bm \rho}) < 0$], the electron effective Hamiltonian reads
\begin{equation} \label{HK}
{\cal H}^K({\bm k}) = \hbar v_0 \bm \sigma \cdot \bm k + {E_g\over 2}\sigma_z \:,
\end{equation}
${\bm k}$ is the electron wave vector measured from the Dirac point, $v_0$ is the Dirac velocity, ${\bm \sigma}$ is the 2D vector consisting of the pseudospin Pauli matrices $\sigma_x, \sigma_y$, and
the hole effective Hamiltonian is related to ${\cal H}^K$ by 
\begin{equation} \label{HK'}
{\cal H}^{h,K'}({\bm k}) = {\cal H}^K(- {\bm k})\:.
\end{equation}

The eigenenergies of the ${\cal H}^K({\bm k})$ are equal to $\pm \varepsilon_{k}$ with $\varepsilon_{k}$ being $\sqrt{(E_g/2)^2+(\hbar v_0 k)^2}$, and the corresponding eigencolumns $u_{\pm,\bm k}$ are
\begin{equation}
\label{u}
 u_{+,\bm k} = \mqty[T_+ e^{-i\varphi_{\bm k}/2}\\ T_-e^{i\varphi_{\bm k}/2}], \quad  u_{-,\bm k} = \mqty[-T_- e^{-i\varphi_{\bm k}/2}\\ T_+e^{i\varphi_{\bm k}/2}],
\end{equation}
where $T_{\pm} = \sqrt{\qty[1\pm E_g/(2\varepsilon_{k})]/2}$, and $\varphi_{\bm k}$ is the azimuthal angle of the vector $\bm k$.

An explicit equation for the matrix element of the excitation of the unbound (but correlated) electron-hole pair can be readily written in terms of the coefficients of the wavefunction expansion in the states of non-interacting particles follows
\begin{equation} \label{Clelh}
{\bm \Psi}_{\rm exc} = \sum_{\lambda_e \lambda_h}\sum_{\bm k} \lambda_h {\cal C}_{\lambda_e\lambda_h}(\bm k) |e, \lambda_e, {\bm k}; h, \lambda_h, -{\bm k} \rangle \:,
\end{equation}
where $\lambda_e = \pm$ and $\lambda_h=\pm$ indicate the one-particle states with positive and negative energies. Note that we assume the normal incidence of the exciting light in which case ${\bm k}_h = - {\bm k}_e$ and the pair momentum $\hbar ({\bm k}_e + {\bm k}_h)$ is zero. In the representation (\ref{Clelh}) the optical absorption matrix element is given by \cite{Leppenen_2020}
\begin{equation}
M({\bm e}) = e v_0 \sum\limits_{\bm k} \left( \text{e}^{{\rm i}\varphi_{\bm k}} e_- R_+ -  \text{e}^{-{\rm i}\varphi_{\bm k}} e_+ R_- \right),
\end{equation}
where ${\bm e}$ is the light polarization vector, $e_\pm=e_x\pm ie_y$, and
\begin{equation}
R_{\pm}({\bm k}) = T_{\pm}^2 {\cal C}^*_{++} + T_{\mp}^2 {\cal C}^*_{--} \mp T_+T_- ({\cal C}^{ *}_{+-} + {\cal C}^{ *}_{-+}).
\end{equation}

The expansion coefficients in Eq.~(\ref{Clelh}) satisfy the following equation 
\begin{equation} \label{eq_for_C}
\sum_{\lambda'_e, \lambda'_h, {\bm k}'} {\cal H}_{\lambda_e, \lambda_h; \lambda'_e, \lambda'_h}({\bm k},{\bm k}') C_{\lambda'_e \lambda'_h}({\bm k}') = E C_{\lambda_e \lambda_h}({\bm k})\:,
\end{equation}
with the effective Bethe-Salpeter two-particle Hamiltonian being
\begin{align}
\label{BS}
& {\cal H}_{\lambda_e, \lambda_h; \lambda'_e, \lambda'_h}({\bm k},{\bm k}') = 	(\lambda_e+\lambda_h)\varepsilon_{k} \delta_{\lambda_e, \lambda'_e} \delta_{\lambda_h, \lambda'_h} \delta_{{\bm k},{\bm k}'}
	\\ &\hspace{1 cm} + {{\cal J}_{\lambda_{e},\lambda_{h};\lambda'_{e}\lambda'_{h}}(\bm k \leftarrow \bm k')}. \nonumber
\end{align}
Here $E=\hbar\omega$ and the kernel is given by
\begin{align}
& 	\hspace{1.5 cm} {\cal J}_{\lambda_{e},\lambda_{h};\lambda'_{e}\lambda'_{h}}(\bm k \leftarrow \bm k') \nonumber\\
&	= V(\abs{\bm k-\bm k'}){(u^\dagger_{\lambda_e,\bm k}u_{\lambda_e',\bm k'})(u^\dagger_{\lambda_h,\bm k}u_{\lambda_h',\bm k'})} 
\end{align}
with $V(|{\bm q}|)$ being the Fourier image of the electron-hole attraction potential
and both $u_{\lambda_e, {\bm k}}$ and $u_{\lambda_h, {\bm k}}$ ($\lambda_e, \lambda_h = \pm 1$) given by Eq.~\eqref{u}.

We compare the results of rigorous calculations with the {\it wide-band gap limit}. 
In this simplified model, only one component of the exciton wavefunction, $\mathcal{C}_{++}$, is taken into account. The three other components are assumed to be zero, and only one equation~\eqref{eq_for_C} (with $\lambda_e=\lambda_h=\lambda'_e=\lambda'_h=+$) is solved in this case where the approximation $2\varepsilon_k \approx E_g+\hbar^2k^2/(2\mu)$ with the exciton reduced mass $\mu=E_g/(4v_0^2)$ is made. This approach is relevant for systems with $E_g$ much larger than the exciton binding energy $E_B$ and for the range $\hbar\omega - E_g \ll E_g$.

Due to the time-inversion symmetry, the optical matrix element for transitions in the $K'$ valley equals to $M^*(\bm e^*)$. As a result, we obtain for the absorption in the right-handed circular polarization ($e_- = \sqrt{2}, e_+=0$) 
\begin{equation} \label{M2}
\sum_{i = K,K'} \abs{M_{i}(\sigma^+)}^2 = 2 (e v_0)^2 \Omega,
\end{equation}
where
\begin{equation}
\label{Osc_strength}
\Omega = \abs{\sum_{\bm k}e^{i\varphi_{\bm k}}R_{+}(\bm k)}^2+\abs{\sum_{\bm k}e^{-i\varphi_{\bm k}}R_{-}(\bm k)}^2.
\end{equation}
Note that since the studied system is nonmagnetic the absorption is independent of the light polarization, and here we use the particular case of circular polarization in order to obtain the result faster.

The absorbance $\eta(\omega)$ is defined as a ratio of the absorption rate to the photon flux. According to the definition of the Sommerfeld factor $S(\omega)$, $\eta(\omega)$ is written as a product
\begin{equation}
\eta(\omega) = \eta_0(\omega) S(\omega),
\end{equation}
where $\eta_0(\omega)$ is the absorbance calculated neglecting the electron-hole interaction. For the  Dirac materials under consideration, one has \cite{eta_0_graphene,eta_0,Kotov_gap,Geim}
\begin{equation} \label{eta_0}
\eta_0(\omega) = {f}{\pi e^2\over 2\hbar c} \qty[1+\qty({E_g\over \hbar \omega})^2]\:.
\end{equation}
For a 2D flake lying on the substrate with the refractive index $n_s$, the factor $f$ is equal to
\[
f = \frac{4}{(n_s + 1)^2}\:.
\]
Note that in the following, for the sake of brevity, we set $f$ to unity.

At $E_g=0$, Eq.~\eqref{eta_0} yields a half of graphene absorbance because graphene has double spin degeneracy. Unlike graphene, the absorbance~\eqref{eta_0} of 2D Dirac material is dependent on the frequency resulting in a factor of two variation as the photon energy $\hbar\omega$ increases from the band gap $E_g$ to the values $\hbar\omega \gg E_g$. 

The interaction between the photocarriers in an extremely thin layer embedded between two thick materials (one of them may be vacuum) occurs via the Rytova--Keldysh potential whose the Fourier-transform is
\begin{equation}
\label{int}
V(q)=-{2\pi e^2\over  q\varkappa(1+q r_0)}.
\end{equation}
Here $\varkappa$ is a half sum of the dielectric constants of the top and bottom media, and 
$r_0$ is the screening radius representing the 2D layer polarizability~\cite{rev2}.
In the case of the interaction~\eqref{int}, the frequency dependence of the absorbance is governed by two dimensionless parameters, namely,
\begin{equation}
g = {e^2\over \varkappa \hbar v_0}, \qquad \tilde{r}_0 =  {E_g r_0 \over 2\hbar v_0 }.
\end{equation}

\begin{figure}[h]
\centering
\includegraphics[width=0.99\linewidth]{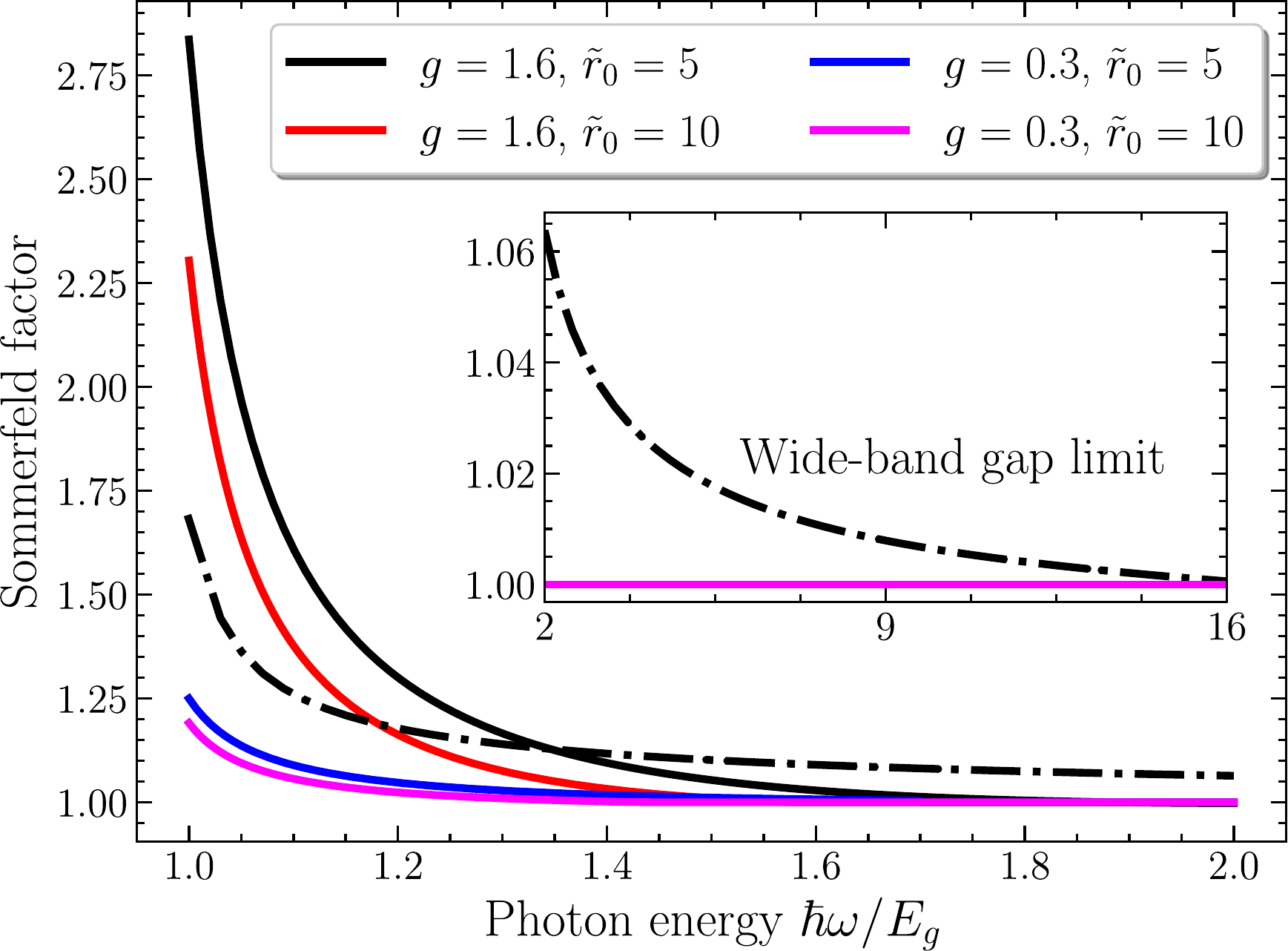}
\caption{Sommerfeld factor frequency dependence for various strengths of the interaction $g$ and the Rytova--Keldysh screening radii $\tilde{r}_0$. The dash-dotted line represents the Sommerfeld factor calculated in the wide-band gap limit with $g = 1.6$ and $\tilde{r}_0 = 5$.
Inset demonstrates the broader range of photon energies.
}
\label{fig:Num_calc}
\end{figure}

We have solved the system of integral equations~\eqref{eq_for_C} for given $g$ and $\tilde{r}_0$ numerically by using the quadrature method \cite{Chuang_par}. 
In Ref.~\cite{Leppenen_2020} we used the dimensionless wave vector ${\bm Q} =a_B {\bm k}$, where $a_B=2\hbar v_0/(g E_g)$ is the effective 2D Bohr radius. In difference to Ref.~\cite{Leppenen_2020}, here we choose for the continuous spectrum the dimensionless variable ${\bm Q} = 2 \hbar v_0{\bm k}/E_g$ independent of the interaction strength $g$.
This choice allows us to find the Sommerfeld factor for a given $g$  via the following working equation
\begin{equation}
S={\Omega(g)\over \Omega(0)}\:, 
\end{equation}
i.e. as a ratio of $\Omega$'s, Eq.~\eqref{Osc_strength}, in the presence and in the absence of interaction.
This ratio does not depend on the normalization procedure, therefore we 
normalized the wavefunction
in the $\bm Q$-space for any value of $g$ similar to the wide-band gap model~\cite{Chuang_cont}. 
The results are presented in Fig.~\ref{fig:Num_calc}. In order to find the absorbance 
one should multiply the found Sommerfeld factor by the $\eta_0(\omega)$ given by Eq.~\eqref{eta_0}.

Figure~\ref{fig:Num_calc} demonstrates the following characteristic features of the Sommerfeld factor in the 2D Dirac systems. The effect of interaction is significant yielding an enhancement 
near the continuous absorbance threshold $\hbar\omega=E_g$. However, the effect sharply falls and becomes negligible already at ${\hbar\omega-E_g \approx 1.6~E_g}$. These are the main qualitative results of the present work.

The dash-dotted line allows us to compare the results of rigorous calculations with the wide-band gap limit which is also treated numerically for the Rytova--Keldysh potential~\eqref{int}.
One can see a remarkable difference of the exact results with this approximation. In the wide-band gap limit, the Sommerfeld factor is underestimated at the threshold and it tends to unity at much higher photon energies, see inset to Fig.~\ref{fig:Num_calc}.

\section{Perturbation theory for weak interaction}
\label{Pert_theor}

\begin{figure}[h]
\includegraphics[width=0.99\linewidth]{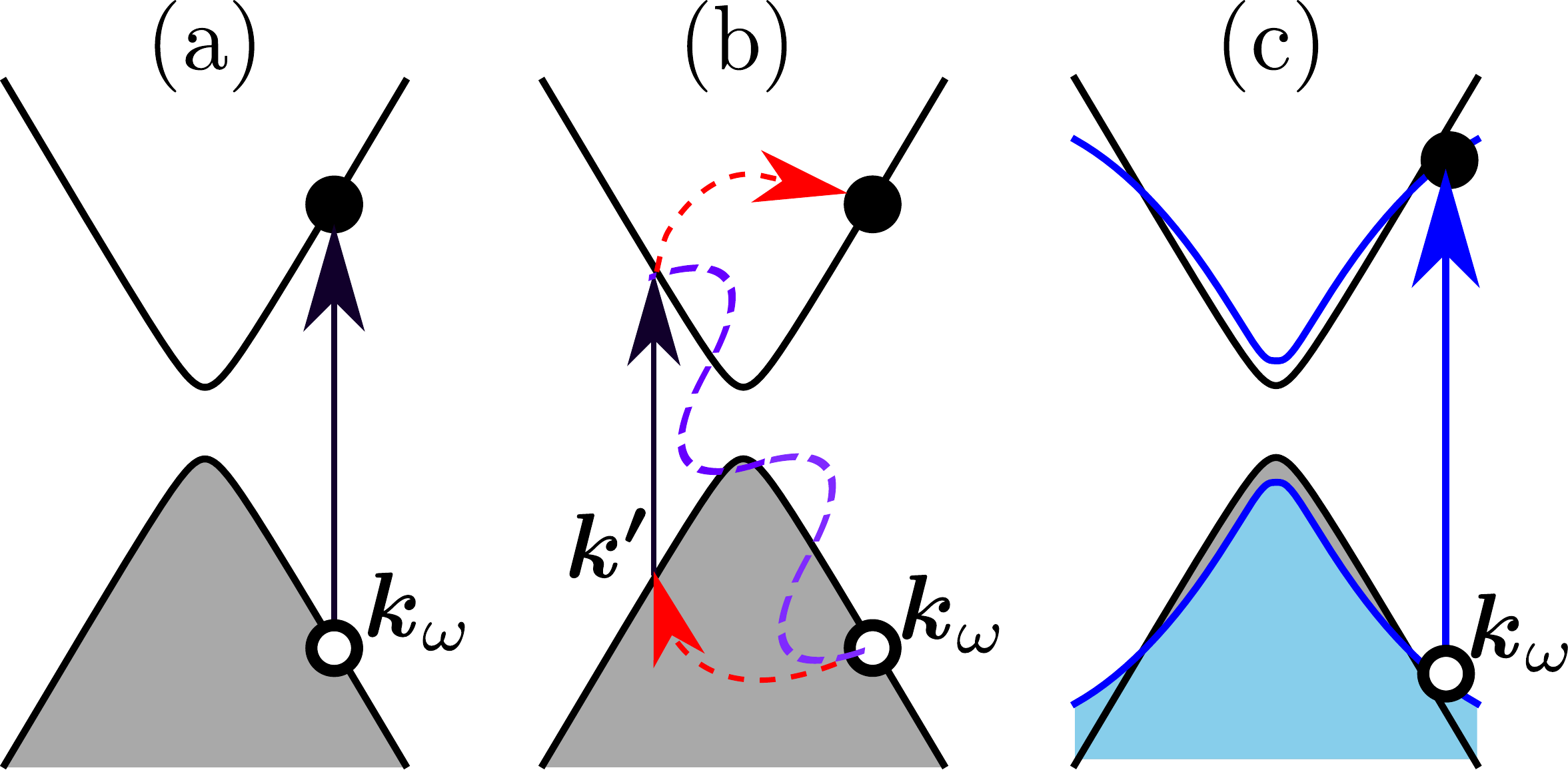}
\caption{Optical transitions contributing to the absorption in the absence of interaction (a) and the first-order corrections. The correction due to the optical transition at $\bm k'$ followed by the simultaneous scattering of the photoelectron $\bm k' \to \bm k_\omega$ and photohole $\bm k_\omega \to \bm k'$ (dashed arrows) caused by the interaction (wavy line) (b) and due to changes of energy dispersion in the conduction and valence bands (c).
The value of $k_\omega$ is defined by Eq.~\eqref{k_omega}.
}
\label{fig:optical_transitions}
\end{figure}

To get physical insight in the results of numerical calculation we derive here the approximated absorbance spectrum taking into account the zeroth and first orders in the interaction strength $g$. It turns out that the main features of the interband continuum-states absorption are already present in the lowest orders. In the perturbation approach, the Sommerfeld factor is presented in the form
\begin{equation}
\label{S_g_lin}
S(\omega) = 1 + g {\mathcal{F}(\omega, E_g, {r}_0)}\:,
\end{equation}
{where ${\cal F}$ is a function independent of $g$.

In wide band gap systems the {$g$-linear} term in Eq. (\ref{S_g_lin}) is obtained as a correction to the optical transition matrix element. The first order scattering} process is illustrated in Fig.~\ref{fig:optical_transitions}(b). However, the single-particle energy spectrum renormalization can also yield a contribution to the Sommerfeld factor of the same order of magnitude. This has been demonstrated for zero-gap materials, graphene~\cite{Mishchenko_2008,Sheehy_2009,Herbut_2010,Kotov_2012,Teber_Kotikov,
Link_2016,Mishchenko_2020} and 3D Weyl semimetals~\cite{Weyl_Jurichich}. This contribution is illustrated in Fig.~\ref{fig:optical_transitions}(c) for a Dirac material with non-zero band gap $E_g$.

The linear correction in Eq.~\eqref{S_g_lin} is conveniently calculated by the diagram technique allowing for determination of the optical conductivity $\sigma(\omega)$
which is related to the absorbance via $\eta = 4\pi\sigma/c$. 
The following diagrams and general equations relating the absorbance $\eta$ with the free-carrier Green's functions are similar to those derived previously for zero-gap materials but the final equations for $\eta$, Eqs. (\ref{vert}) and  (\ref{self}), depend explicitly on the band gap $E_g$ and are original.

\begin{figure}[h]
\includegraphics[width=0.99\linewidth]{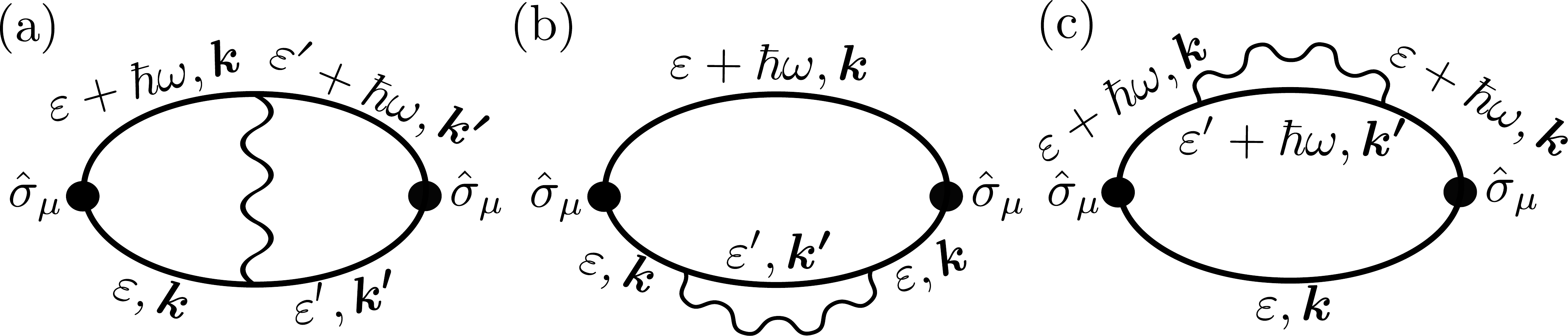}  
		\caption{The diagrams yielding the linear in $g$ correction to the optical conductivity and Sommerfeld factor \eqref{S_g_lin}. Both the vertex  {(a)} and the self energy (b,c) renormalizations contribute to the Sommerfeld factor.}
\label{fig:diagrams}
\end{figure}

Thus, the correction to the optical conductivity is given by a sum of the vertex and self-energy contributions
shown by three diagrams in Fig.~\ref{fig:diagrams}. The vertex renormalization, Fig.~\ref{fig:diagrams}(a), corresponds to the process depicted in Fig.~\ref{fig:optical_transitions}(b), and the self-energy contributions, Figs.~\ref{fig:diagrams}(b) and~(c), correspond to the renormalization of the energy dispersion in the valence and conduction bands illustrated in Fig.~\ref{fig:optical_transitions}(c). Needless to say, the interaction-free absorption process shown in Fig.~\ref{fig:optical_transitions}(a) is described by the pure bubble diagram.
{The energy gap renormalization has been calculated earlier~\cite{Kotov_gap} but its effect on the optical conductivity has not been analyzed.}

In Fig.~\ref{fig:diagrams}, the electric current component operator $\hat j_\mu$ in each vertex equals to $e v_0 \hat\sigma_\mu$ ($\mu=x,y$), and the single-particle Green functions are the second-rank matrices~\cite{Mishchenko_2008}
\begin{equation}
\hat{G}_{\varepsilon}(\bm k)= \frac{\hat{P}_+(\bm k)}{\varepsilon-\varepsilon_k+{\rm i}0}+\frac{\hat{P}_-(\bm k)}{\varepsilon+\varepsilon_k- {\rm i} 0}, 
\end{equation}
where the projection operators onto the conduction- and valence-band states for the gapped system are given by  
\begin{equation}
\label{proj}
\hspace{-0.3 cm} \hat{P}_{\pm}=\frac{\varepsilon_k \pm {\cal H}^K({\bm k})}{2 \varepsilon_k}=\mqty[T_\pm^2 & \pm T_+T_-e^{-i\varphi_{\bm k}} \\ \pm T_- T_+e^{i\varphi_{\bm k}} &  T_\mp^2]
\end{equation}
with $T_\pm$ from Eq.~\eqref{u}. The sum of three diagrams is reduced to
\begin{equation}
	\eta-\eta_0 =  {(ev_0)^2 \over i  \pi c \omega}\sum_{\bm k, \bm k'}\int \int \dd{\varepsilon}\dd{\varepsilon}' V({\bm k-\bm k'}) (\Phi_\text{vert}+\Phi_\text{self}),
\end{equation}
where~\cite{Link_2016}
\begin{align}
&\Phi_\text{vert}=\sum_{\mu = x,y}\Tr[\hat{\sigma}_\mu \hat{G}_{\varepsilon'}(\bm k')\hat{G}_{\varepsilon}(\bm k)\hat{\sigma}_\mu\hat{G}_{\varepsilon+\hbar \omega}(\bm k)\hat{G}_{\varepsilon'+\hbar \omega}(\bm k')], \nonumber
\\
&\Phi_\text{self}= 2\sum_{\mu = x,y}\Tr[\hat{\sigma}_\mu \hat{G}_{\varepsilon+\hbar \omega}(\bm k) \hat{\sigma}_\mu \hat{G}_{\varepsilon}(\bm k) \hat{G}_{\varepsilon'}(\bm k') \hat{G}_{\varepsilon}(\bm k)]. \nonumber
\end{align}
Here we introduced half-sums of the optical conductivities for two linear polarizations along the axes $\mu=x,y$ using an independence of the absorbance on the polarization. The factor of 2 in $\Phi_\text{self}$ accounts for equal contributions of two self-energy diagrams in Figs.~\ref{fig:diagrams}(b) and~(c), i.e. the energy spectrum renormalization in both the conduction and valence bands. At the end the total result was multiplied by 2 for the account of the $K'$ valley contribution.

Calculation shows that the vertex correction has the form
\begin{widetext}
\begin{equation}
\label{vert}
\eta_\text{vert}(\omega)= - \frac{\pi e^2}{\hbar c}\pv \sum_{\bm k'} V(|{\bm k}_{\omega} - {\bm k'}|)
\frac{E_g^2[E_g^2+3(\hbar \omega)^2+4 (\hbar k' v_0)^2]+ (2 \hbar v_0)^2 k_\omega k' \cos\theta \qty[E_g^2+(\hbar \omega)^2+ (2 \hbar v_0)^2 k_\omega k' \cos\theta]}{(\hbar \omega)^2 \sqrt{E_g^2+4 (\hbar k' v_0)^2} (2\hbar  v_0)^2(k'^2-k_\omega^2)}, 
\end{equation}
where $\pv$ stands for the Cauchy principle value {and we introduced $\bm k_\omega $ as a vector of arbitrary direction and the absolute value fixed by the energy conservation law
\begin{equation}
\label{k_omega}
k_\omega = {\sqrt{(\hbar \omega)^2-E_g^2} \over 2\hbar v_0},
\end{equation}
and $\theta$ is the angle between the vectors $\bm k'$ and $\bm k_\omega$. 
The contribution of two self-energy diagrams can be written in the form}
\begin{equation}
\label{self}
\eta_\text{self}(\omega) = - \frac{\pi e^2}{2 \hbar^2 c \omega}\pdv{\hbar \omega}\qty{\qty[1+\qty(E_g\over \hbar \omega)^2]\qty[(2\hbar k_\omega)^2v_0\delta v(k_\omega)+E_g \delta E_g(k_\omega)]}-\frac{\pi e^2}{\hbar c}\delta E_g(0)\delta(\hbar\omega-E_g) .
\end{equation}
\end{widetext}
Deriving Eq.~\eqref{self} we used the relation 
\[
{1\over (2\varepsilon_k-\hbar\omega -i0)^2} 
= \pdv{\hbar \omega}\qty[\pv{1\over 2\varepsilon_k-\hbar\omega } + i\pi\delta(2\varepsilon_k-\hbar\omega)]
\]
and the interaction-induced renormalization of the velocity and the energy gap given by~\cite{Kotov_2012,Kotov_gap}
\begin{equation} \label{deltav}
	\delta v(k) = -\frac{v_0}{k}\sum_{\bm k'}\frac{k'\cos\theta V(\abs{\bm k-\bm k'})}{\sqrt{E_g^2+(2\hbar v_0 k')^2}},
\end{equation}
\begin{equation}
\label{d_E_g}
	\delta E_g(k) = -E_g\sum_{\bm k'}\frac{V(\abs{\bm k-\bm k'})}{\sqrt{E_g^2+(2\hbar v_0 k')^2}} .
\end{equation}
The term in Eq.~\eqref{self} proportional to the $\delta$-function represents a blueshift of the absorption threshold due to the energy gap renormalization. Below we consider only the range $\hbar \omega>E_g$ and neglect this term. 

The obtained equations~\eqref{vert} and~\eqref{self} are valid for any weak interaction potential $V(q)$. Below we consider first the general case of the Rytova--Keldysh potential and then the pure Coulomb potential ($r_0=0$).

\begin{figure}[h]
\includegraphics[width=0.99\linewidth]{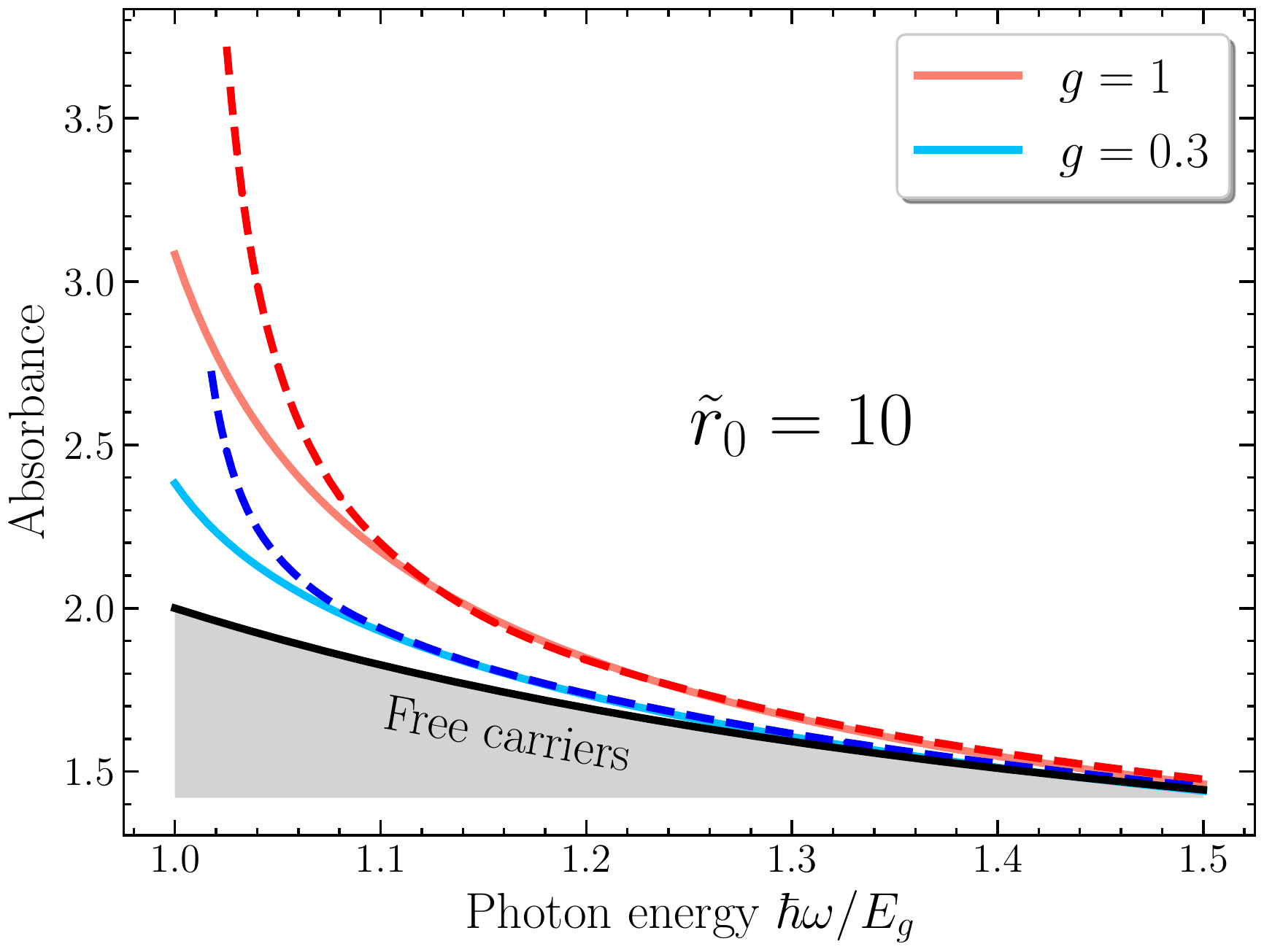}
\caption{Comparison of the absorbance calculated exactly (solid lines) and perturbatively (dashed lines) for the Rytova--Keldysh interaction potential~\eqref{int} with $\tilde{r}_0= 10$. The black solid line shows the free carrier absorbance $\eta_0(\omega)$, Eq.~\eqref{eta_0}.
The absorbance is given in units of $\pi e^2/( 2\hbar c)$.}
\label{fig:Pert_theory_compar}
\end{figure}

We have calculated the absorbance $\eta(\omega)$ by numerical evaluation of Eqs.~\eqref{vert} and~\eqref{self} which converge for the Rytova--Keldysh interaction~\eqref{int} with  ${r}_0 \neq 0$. The {results are} shown in Fig.~\ref{fig:Pert_theory_compar} where they are compared with the exact {calculation}. The photon energy where the perturbative and exact results start to match is well described by the relation ${\hbar\omega-E_g \approx E_B}$, where $E_B$ is the binding energy of the 2D exciton obtained in the Dirac model for the Rytova--Keldysh potential~\cite{Leppenen_2020}. In particular, for {higher values of $g$ the match border shifts to} higher photon energies, see Fig.~\ref{fig:Pert_theory_compar}.

One can see from Fig.~\ref{fig:Pert_theory_compar} that the two main features of the {absorption} spectrum, namely, a strong enhancement of the near band-gap absorbance and rapid decay of the interaction effect with {the increasing light frequency}, are captured by the perturbative approach.

\subsection{Graphene with the Rytova--Keldysh interaction}

The first-order correction to the optical conductivity of graphene has been studied for various electron-hole attractive potentials including unscreened and screened Coulomb interaction as well as the short-range interaction~\cite{Mishchenko_2008,Sheehy_2009,Herbut_2010,Kotov_2012,Teber_Kotikov,
Link_2016,Mishchenko_2020}. In order to obtain finite results, the cut-off of the potential is needed in all these cases. Here we consider the Rytova--Keldysh potential~\eqref{int} in which case the $q^{-2}$ decrease of the potential at large $q$ ensures the convergence of the sums in Eqs.~(\ref{vert}) and~(\ref{self}), and the  regularization is not required. 

For graphene with $E_g=0$, Eqs.~\eqref{vert}  and~\eqref{self} are simplified to 
\begin{align}
\label{self_graphene}
&	\eta_{\rm self} = - g\frac{e^2}{8\hbar c}\frac{1}{w}\pdv{w}\qty[w^2 \int\limits_0^{2\pi} \dd{\theta} \int\limits_0^{\infty} \dd{x} \frac{1-x\cos\theta}{(1+wx)u}],
\\
\label{vert_graphene}
&\eta_{\rm vert} = g\frac{e^2}{4 \hbar c}\int\limits_{0}^{2\pi} \dd{\theta} \: \pv \int\limits_{0}^{\infty} \dd{x} \frac{x\cos\theta (1+x\cos\theta)}{u(1+wu)(x^2-1)}\:,
\end{align}
where $u=\sqrt{x^2-2x\cos\theta+1}$
and
\begin{equation}
\label{w}
w={r_0\omega \over 2v_0}.
\end{equation}
The dimensionless first-order correction $\mathcal{F}(\omega, E_g, {r}_0)=(\eta/\eta_0-1)/g$ depends at $E_g=0$ on the single dimensionless parameter~(\ref{w}) which can be constructed from $\omega$, $v_0$ and $r_0$. 

\begin{figure}[h]
	\centering
	\includegraphics[scale=0.45]{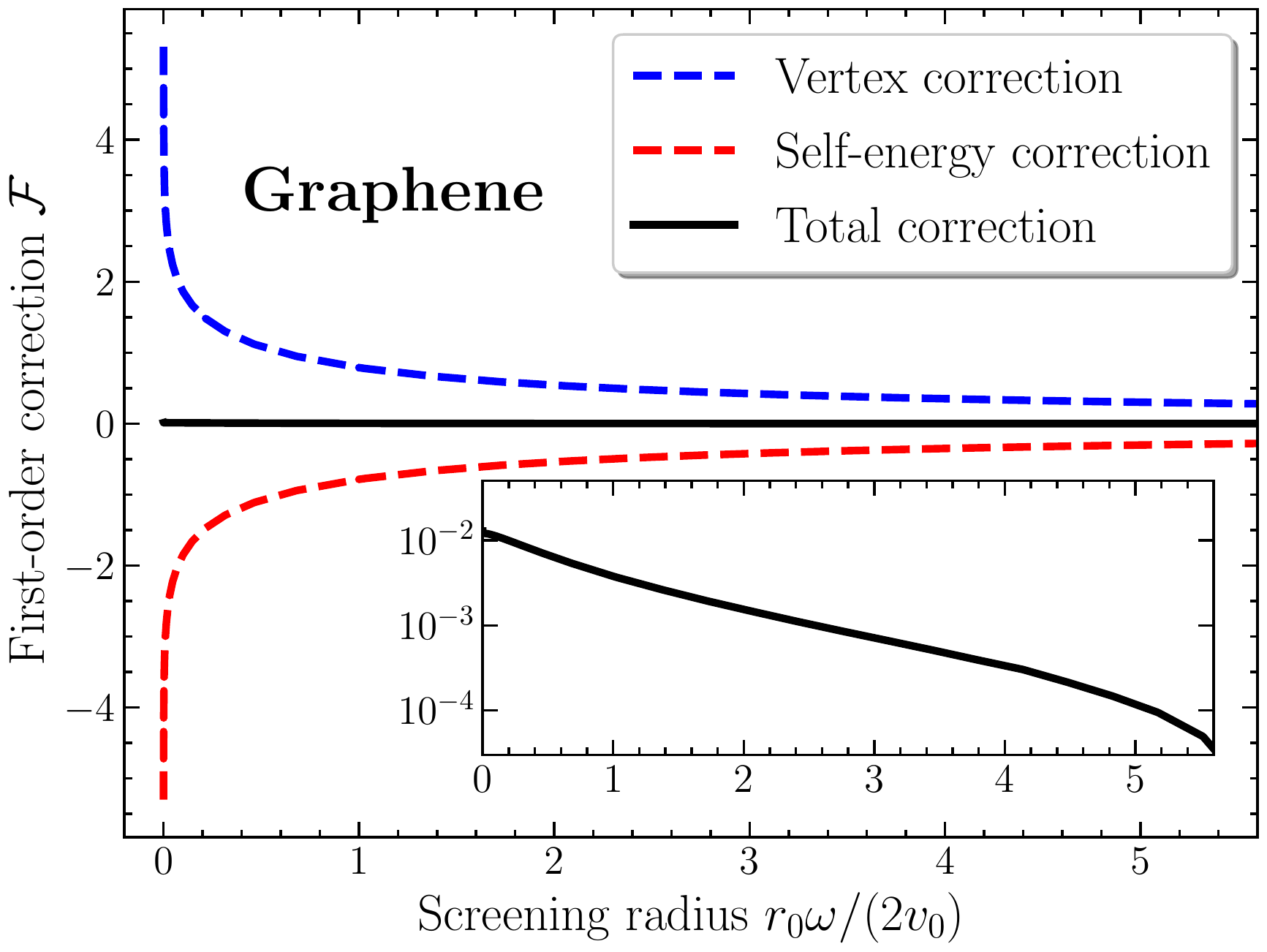} 
	\caption{ The $w$ dependence of contributions to $\mathcal{F}$ due to the vertex diagrams and self-energy (dashed lines),	and their sum (black line) in graphene  with the Rytova--Keldysh interaction potential. Inset shows the sum in the larger scale. }
	\label{fig:RK-graphene}
\end{figure}

The dependence of $\mathcal{F}$ on $w$ calculated by Eqs.~\eqref{self_graphene} and~\eqref{vert_graphene} is plotted in Fig.~\ref{fig:RK-graphene}. 
The contributions $\eta_{\rm vert}(w)$ and $\eta_{\rm self}(w)$ are very close to each other by absolute value and almost completely compensate each other. As a result, the sum is smaller by 2-4 orders compared to each separate term.

The Sommerfeld factor in graphene demonstrates a specific property of the linear energy spectrum:  the absorbance enhancement is almost absent. For the pure Coulomb potential it follows from the fact that only one energy scale, $\hbar \omega$, is present in the problem. Therefore, the dimensionless Sommerfeld factor should be constant in the whole  frequency range, and, since at high frequencies  the Coulomb-induced electron-hole correlation almost vanishes, this constant is close to unity~\cite{Mishchenko_2008,Sheehy_2009,Teber_Kotikov}. This means that the correction $\mathcal{F}(E_g=0)$ is very small. For the Rytova--Keldysh potential, a new energy scale, $\hbar v_0/r_0$, emerges and, thereby, the Sommerfeld factor gets dependent on the relation between frequency $\omega$ and the ratio $r_0/v_0$, Eq.~\eqref{w}. This dependence is shown in the inset in Fig.~\ref{fig:RK-graphene}. However, as compared to the pure Coulomb potential, the screened potential~\eqref{int} should result in a smaller Sommerfeld factor. Therefore $\mathcal{F}(w)$ is a decreasing function going from $\approx 0.013$ at $w \to 0$ to zero at large~$w$. We thus conclude that in graphene the Sommerfeld factor is close to unity at any degree of screening.

\subsection{Wide-band gap limit}

In the wide-band gap limit $E_B \ll E_g \lesssim \hbar\omega$
and for long-range Rytova--Keldysh potential, 
\begin{equation} \label{longrange}
r_0 \gg \hbar v_0/E_g\:,
\end{equation}
one can check that the main contribution to $\eta_\text{vert}(\omega)$ comes from $k' \sim k_\omega  \ll E_g/(\hbar v_0)$. 
In Fig.~\ref{fig:optical_transitions}(b) it corresponds to the situation where both $\bm k_\omega$ and $\bm k'$ are close to the valence band top. In this case, neglecting the small values of $\hbar v_0 k'/E_g$ and $\hbar v_0 k_\omega/E_g$ in Eq.~\eqref{vert}, we obtain an analytical result
\begin{equation}
\label{eta_vert_wbg}
{\eta_\text{vert}^\text{(wbg)} \over \eta_0} = g \frac{\pi}{2\sqrt{2}}\sqrt{\frac{E_g}{\hbar \omega-E_g}}\qty[1-\frac{2z \arccos (z^{-1})}{\pi\sqrt{z^2-1}}],
\end{equation}
where $\eta_0= \eta_0(E_g/\hbar) = \pi e^2/(\hbar c)$ and ${z = 2 \sqrt{2} \tilde{r}_0\sqrt{\hbar \omega/E_g-1}}$. For $z<1$ the expression in brackets
can be conveniently rewritten as
\[
1-\frac{2z}{\pi\sqrt{1-z^2}}\ln(\frac{1+\sqrt{1-z^2}}{z}).
\]

The self-energy correction to the Sommerfeld factor in the wide-band gap limit is negligible as compared to $\eta^{\rm (wbg)}_{\rm vert}$ given by Eq.~(\ref{eta_vert_wbg}). This can be demonstrated taking into account that in Eqs. (\ref{deltav}) and (\ref{d_E_g}) the upper limit in the summation over ${\bm k}'$ can be set $k' = E_g/\hbar v_0$.

\subsection{Gapped Dirac material}

Now we turn to the gapped Dirac materials and introduce the vertex and self-energy contributions to the first-order correction $\mathcal{F}(\omega,E_g,r_0)$ according to
\begin{equation}
\label{F_self_vert}
 \mathcal{F}_\text{vert} = \frac{\eta_\text{vert}}{g \eta_0}\:,\quad \mathcal{F}_\text{self} = \frac{\eta_\text{self}}{g \eta_0}\:.
\end{equation}

\begin{figure}[h]
\includegraphics[width=0.99\linewidth]{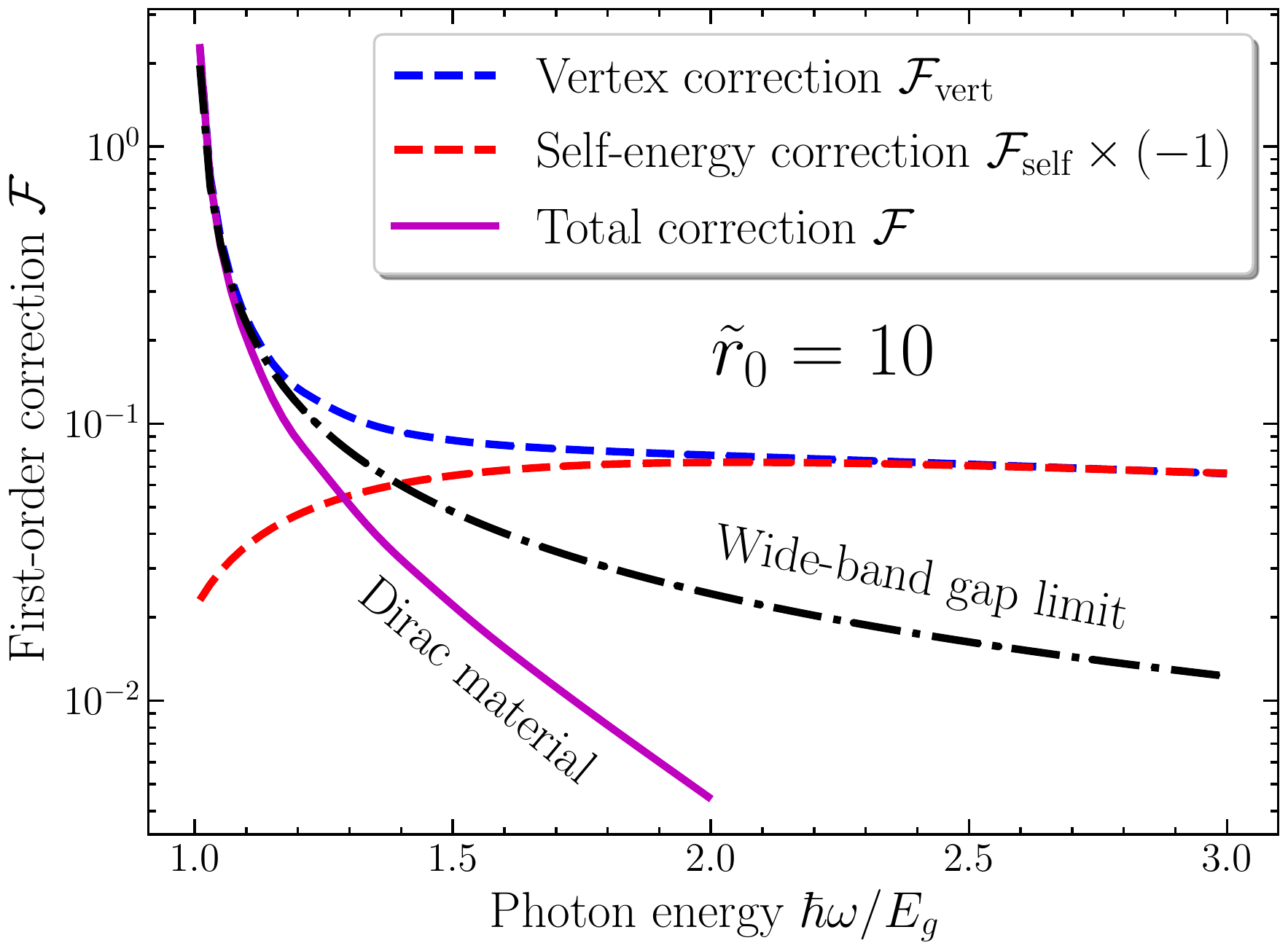}
\caption{First-order correction to the Sommerfeld factor $\mathcal{F}=(S-1)/g$, 
Eq.~\eqref{S_g_lin}, calculated for $\tilde{r}_0 = 10$ (solid line). Dashed lines are the vertex and self-energy contributions~\eqref{F_self_vert}. The dash-dotted curve is the correction in the wide-band gap limit, Eq.~\eqref{eta_vert_wbg}.}
\label{fig:Two_diagr_a}
\end{figure}

In Fig.~\ref{fig:Two_diagr_a} we compare the contributions $\mathcal{F}_\text{vert}$, $\mathcal{F}_\text{self}$ and their sum $\mathcal{F}_\text{vert} + \mathcal{F}_\text{self}$, calculated by Eqs.~\eqref{vert} and~\eqref{self} for arbitrary ratio $\hbar\omega/E_g > 1$, with the result of wide-band gap approximation. The latter exceeds the exact result at high photon energies but works well near the absorption threshold at $\hbar\omega -E_g \lesssim 0.15~E_g$. This is expected because the range of applicability of the wide-band gap approximation is $\hbar\omega-E_g \ll E_g$. 
A bit smaller value of the correction at the absorption threshold obtained in the wide-band gap approximation is caused by the assumption of small $k'$ in Eq.~\eqref{vert} which leads to underestimation of the result. This difference is caused by the short-range nature of the screened interaction~\eqref{int} even at $\tilde{r}_0=10$.

The both major features of the Sommerfeld factor in 2D Dirac systems revealed in Fig.~\ref{fig:Num_calc} are also demonstrated in Fig.~\ref{fig:Two_diagr_a}. One can see that each of the two contributions,  the vertex and self-energy ones, is substantial. The contributions have opposite signs and compensate each other to some extent in the range $\hbar \omega/E_g < 1.3$, to a large degree for $1.7 > \hbar \omega/E_g > 1.5$ and strongly at $\hbar \omega/E_g > 2$. Thus, the effect of interaction on the absorbance is weak far from the threshold. This is expected because, at $\hbar\omega \gg E_g$, the presence of the gap is not significant, and gapped systems behave as graphene, where the two contributions almost completely compensate each other and the Sommerfeld factor is very close to unity, see Fig.~\ref{fig:RK-graphene}.

\begin{figure}[h]
\includegraphics[width=0.99\linewidth]{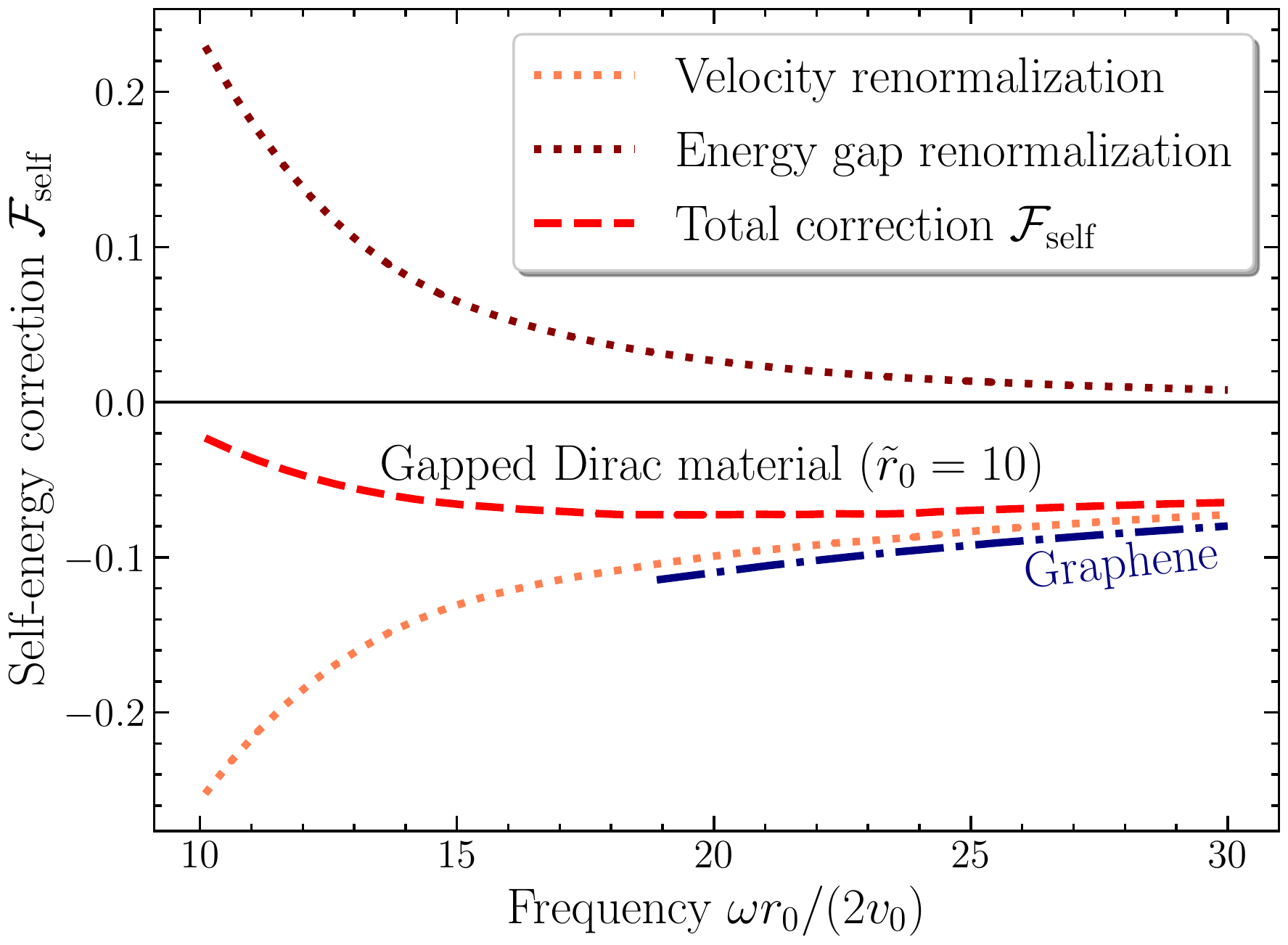}
\caption{Decomposition of the total self-energy correction $\mathcal{F}_\text{self}$ into two contributions due to the velocity and energy gap renormalizations $\delta v(k_\omega)$ and $\delta E_g(k_\omega)$ in Eq.~\eqref{self}, respectively.  The lines are calculated for $\tilde{r}_0 = r_0E_g/(2 \hbar v_0)=10$. Dash-dotted line shows the self-energy correction in graphene.}
\label{fig:Two_diagr_b}
\end{figure}

The difference of gapped materials from graphene in the frequency range near $E_g$ is especially clear from Fig.~\ref{fig:Two_diagr_b}. In fact, the self-energy correction consists of two terms in Eq.~\eqref{self} determined by the renormalizations $\delta v(k_\omega)$ and $\delta E_g(k_\omega)$. In graphene, the energy gap and hence its renormalization are absent, and only the velocity renormalization contributes to the absorbance, with the net contribution being negative~\cite{Mishchenko_2008}. In Fig.~\ref{fig:Two_diagr_b} the frequency dependence of the total correction is shown by dashed line. In a gapped material the interaction widens the gap $E_g$ and, according to Eq.~\eqref{eta_0}, leads to an enhancement of the absorbance. This affects the correction near the band gap almost compensating the correction from the velocity renormalization. However, the latter is stronger, therefore the total self-energy correction is negative at any frequency, Figs.~\ref{fig:Two_diagr_a} and~\ref{fig:Two_diagr_b}. At high frequencies a presence of the gap becomes unimportant, and the self-energy correction coincides with that in graphene.

\subsection{Coulomb interaction potential}

For the 2D Coulomb potential~(\ref{int}) with ${r}_0=0$, both the vertex and self-energy corrections diverge logarithmically. The situation is similar to the case of graphene ($E_g=0$) where these divergences have been investigated in detail~\cite{Mishchenko_2008,Sheehy_2009,Herbut_2010,Kotov_2012,Teber_Kotikov}. The correct way to regularize the problem preserving the Ward identity is to use the cut-off Coulomb potential
\begin{equation}
\label{cutoff}
	V_c(q) = -\frac{2 \pi e^2}{{\varkappa} q}\Theta(\Lambda-q),
\end{equation}
where $\Theta(x)$ is the Heaviside function, and $\Lambda$ is the {cut-off wave vector}.
Then we get, with logarithmic accuracy,
\begin{eqnarray} \label{etaCoul}
&&	\eta_\text{vert, Coul} = \eta_0(\omega) {g\over 2} \ln\qty(8 {\hbar} v_0 \Lambda \over \sqrt{(\hbar \omega)^2-E_g^2}),\\
&&\eta_\text{self, Coul} = \qty[ \qty(E_g\over \hbar\omega)^2 -1]\eta_{\text{vert, Coul}}.
\end{eqnarray}
As a result, {the correction to the Sommerfeld factor (\ref{S_g_lin}) takes the form}
\begin{eqnarray}
&&S(\omega) - 1 = \frac{\eta - \eta_0}{\eta_0} = g \mathcal{F} (\omega, E_g,0)  \nonumber \\
&&={g\over 2}\qty(E_g\over \hbar \omega)^{2} \ln\qty(8 {\hbar} v_{0}\Lambda \over \sqrt{(\hbar\omega)^2-E_g^{2}})+ g{\cal D} \qty(E_g\over \hbar \omega), \label{F_Coulomb} 
\end{eqnarray}
where ${\cal D}(x)$ is a smooth function. While deriving Eqs.~(\ref{etaCoul})-(\ref{F_Coulomb}) we assumed $(\hbar\omega - E_g)/E_g \gg g$, and within this assumption the factor $S(\omega)$ is finite. 

For the case of graphene, i.e. at $E_g=0$, the logarithmic contribution cancels while the function ${\cal D}$ reduces to a constant ${\cal D}(0)={(19-6\pi)/12\approx 0.0125}$~\cite{Mishchenko_2008,Sheehy_2009,Teber_Kotikov}. In the gapped Dirac materials, the energy $\hbar v_0 \Lambda$ exceeds by far the bang gap and the logarithmic term in Eq.~(\ref{F_Coulomb}) prevails over the smooth term. However,  this large value decreases sharply with increasing the photon energy due to the square root denominator and the inverse quadratic dependence, $\omega^{-2}$, of the prefactor. 

In the opposite limit of a wide-band gap 2D semiconductor, for the pure Coulomb potential we get from Eq.~\eqref{eta_vert_wbg} setting ${z \to 0}$:
\begin{equation}
\label{FparC}
{\eta_\text{Coul}^\text{(wbg)} \over \eta_0}= g\frac{\pi}{2\sqrt{2}}\sqrt{\frac{E_g}{\hbar \omega-E_g}}\:.
\end{equation}
Since Eq.~(\ref{eta_vert_wbg}) is derived under the condition (\ref{longrange}) the small values of $z$ presuppose the frequency range close to the band edge, $\hbar \omega - E_g \ll E_g$. Beyond this range the Sommerfeld factor can be better described by Eq.~(\ref{F_Coulomb}).

The correction (\ref{FparC}) can be also obtained by the expansion up to the linear order in $g$ of the well-known analytical result for the 2D Sommerfeld factor in the band model with the parabolic free-carrier dispersion and the Coulomb potential~\cite{Sugano1966,Miller1984,Haug_Koch_book,EL_book,EL_book} 
\begin{equation}\label{Som_par}
	S_\text{Coul}^\text{(wbg)}(\omega) = \frac{2}{1+\exp\qty(-\pi \sqrt{\frac{E_B}{\hbar \omega-E_g}})},
\end{equation}
where the exciton binding energy in the parabolic model is given by $E_B= g^2 E_g/2$~\cite{Leppenen_2020}. 

It is worth noting that the enhancement of the absorbance at the threshold for the  Coulomb interaction in Eq.~(\ref{Som_par}) is just a factor of 2. Our calculation shows that the enhancement of $\eta$ in the 2D Dirac materials is stronger even for the screened Rytova--Keldysh potential, see Figs.~\ref{fig:Num_calc} and~\ref{fig:Pert_theory_compar}.

\begin{figure}[h]
	\includegraphics[scale=0.49]{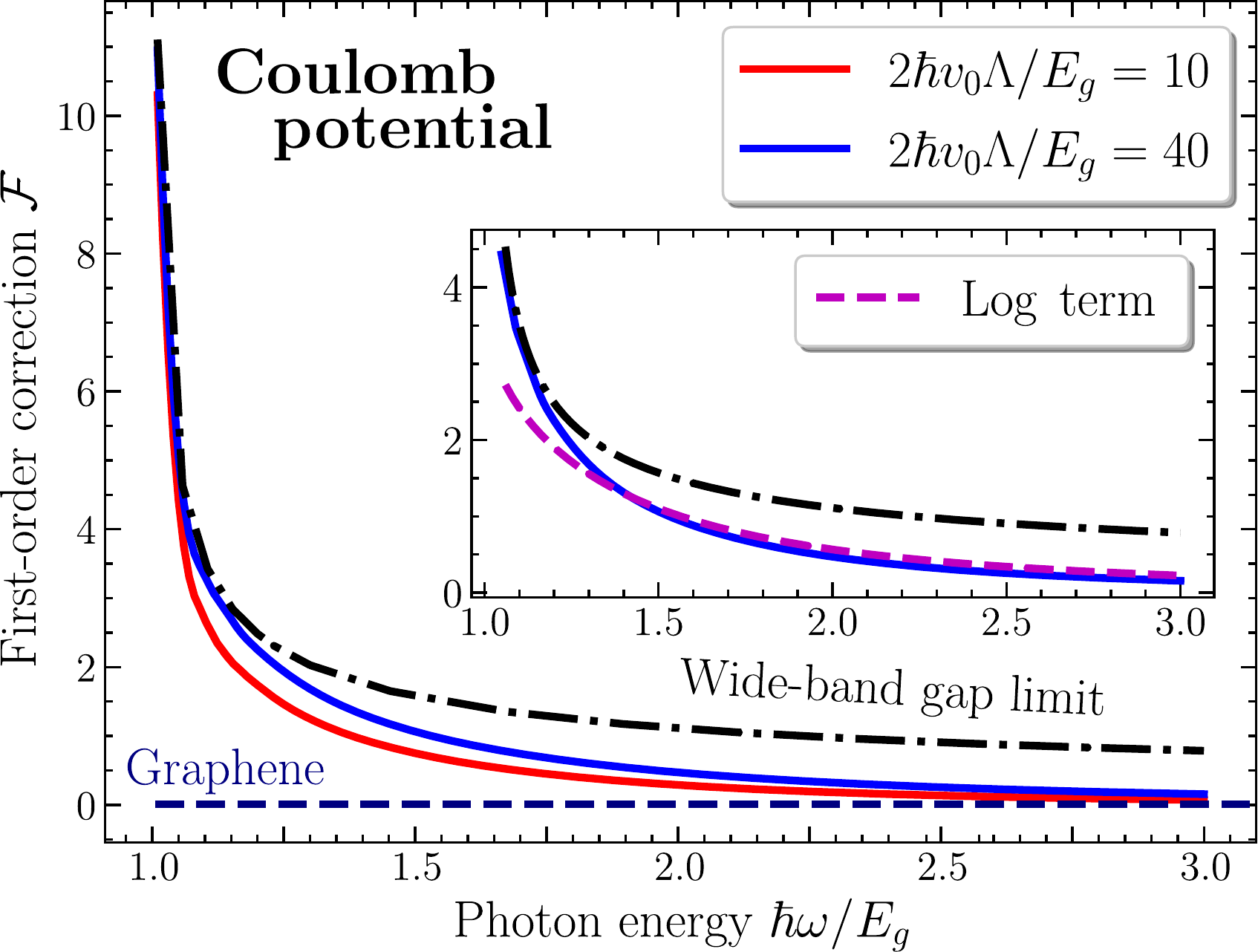} 
	\caption{The first-order correction $\mathcal{F}$ for  the cut-off Coulomb potential~\eqref{cutoff}. Calculations are performed by Eqs.~\eqref{vert},~\eqref{self} (solid lines) and in the wide-band gap limit by Eq.~\eqref{FparC} (dash-dotted line). {In the chosen scale the graphene value $\mathcal{F}=0.0125$ shown by the dashed line lies within the line width.} Inset: the same curves with the logarithmic term~\eqref{F_Coulomb} shown by dashed line for $2\hbar v_0 \Lambda/E_g=40$. }
	\label{fig:CoulLamb}
\end{figure}

In Fig.~\ref{fig:CoulLamb}, we compare the numerical calculation of the first-order correction $\mathcal{F}$ with the gapless graphene value $\mathcal{F}=0.0125$ and the opposite, wide-band gap, limit given by Eq.~\eqref{FparC}. 
The exact result interpolates between these two limiting cases which take place at $\hbar\omega/E_g \gg 1$ and $\hbar\omega/E_g \gtrsim  1$, respectively. Inset to Fig.~\ref{fig:CoulLamb} demonstrates that the pure logarithmic contribution, the first term in Eq.~\eqref{F_Coulomb}, matches well with the exact result already at $\hbar\omega \gtrsim 1.3~E_g$. We see that the wide-band gap expression~\eqref{FparC} and the logarithmic dependence~\eqref{F_Coulomb} jointly describe the first-order correction to the Sommerfeld factor on the whole frequency range. 

\section{Conclusion}
\label{Concl}

In this paper, we have studied the frequency dependence of the Sommerfeld factor in 2D Dirac materials where the electron-hole interaction energy is not too small as compared to the energy gap.
In this case the two-particle wave function is described by four components with allowance for each particle, an electron and a hole, to occupy both the conduction and valence band states. Both the Rytova--Keldysh and pure Coulomb interaction potentials have been used in the derived theory. The theory takes into account not only the electron-hole scattering but also single-particle self-energy renormalization. We have found that in the materials under consideration, in great contrast with the wide band gap nanosystems,  the interaction-induced enhancement of the above-edge light absorption is very strong near the edge and falls sharply with the increasing light frequency. These two features of the Sommerfeld factor behaviour are well captured by the first-order approximation of weak electron-hole interaction. The analysis shows that an absence of the interaction-induced enhancement of the absorbance at high frequencies is caused by a compensation of the vertex and self-energy contributions.
\acknowledgments

N.V.L. and E.L.I. acknowledge the financial support of the Russian Science Foundation (Project~19-12-00051). The work of N.V.L. and L.E.G. was supported by the Foundation for the Advancement of Theoretical Physics and Mathematics ``BASIS''.

\end{document}